# FOCUSING TESTING BY USING INSPECTION AND PRODUCT METRICS

FRANK ELBERZHAGER, STEPHAN KREMER

*Fraunhofer IESE,
Kaiserslautern, Germany
{frank.elberzhager, stephan.kremer}@iese.fraunhofer.de*

JÜRGEN MÜNCH

*University of Helsinki,
Helsinki, Finland
juergen.muench@cs.helsinki.fi*

DANILO ASSMANN

*Vector Informatik GmbH
Stuttgart, Germany
danilo.assmann@vector.com*

A well-known approach for identifying defect-prone parts of software in order to focus testing is to use different kinds of product metrics such as size or complexity. Although this approach has been evaluated in many contexts, the question remains if there are further opportunities to improve test focusing. One idea is to identify other types of information that may indicate the location of defect-prone software parts. Data from software inspections, in particular, appear to be promising. This kind of data might already lead to software parts that have inherent difficulties or programming challenges, and in consequence might be defect-prone. This article first explains how inspection and product metrics can be used to focus testing activities. Second, we compare selected product and inspection metrics commonly used to predict defect-prone parts (e.g., size and complexity metrics, inspection defect content metrics, and defect density metrics). Based on initial experience from two case studies performed in different environments, the suitability of different metrics for predicting defect-prone parts is illustrated. The studies revealed that inspection defect data seems to be a suitable predictor, and a combination of certain inspection and product metrics led to the best prioritizations in our contexts. In addition, qualitative experience is presented, which substantiates the expected benefit of using inspection results to optimize testing.

## 1. Introduction

Software and software systems, such as mobile phones, cars, or medical devices, are part of everyone's life. These systems are continuously increasing in size and complexity, which also increases the risk of failures that might lead to serious consequences, such as high rework costs or loss of reputation for the software-developing company. In order to develop high-quality software products, a large number of different analytic quality assurance techniques exist, such as various inspection and testing techniques. However, such quality assurance activities sometimes consume up to 50% of the overall development effort [1]. Consequently, new approaches are needed that can cope with these challenges.

*Frank Elberzhager, Stephan Kremer, Jurgen Munch, Danilo Assmann*

Different strategies have been developed in the past to make quality assurance, and especially testing, more efficient [19]. One example is automation. The idea is to automate tasks that consume much more effort when done manually, such as test case generation or the analysis of test results. Another strategy is to predict defect-proneness based on certain metrics and context-specific data. For example, assuming that complex parts are more defect-prone, cyclomatic complexity can be calculated, and the most complex parts can be tested more intensively. Such predictions can be used to focus quality assurance activities on certain parts and thus to support allocating efforts more sufficiently. However, results from early quality assurance activities, such as inspections or reviews, are usually not considered when focusing testing, even though one of the goals of inspections is the same as for testing, namely to find defects.

Consequently, we suggest stronger integration of inspections and testing activities in order to exploit synergy effects, respectively to address the above-mentioned challenges. Today, if both inspections and testing are conducted, they are typically applied in isolation without any information exchange between them. With stronger integration, tests can be focused better based on the inspection results, leading to more defect detections or less testing effort [4]. Thus, we propose the integrated inspection and testing approach In$^2$Test, which is explicitly capable of focusing testing activities on certain parts of a system or on certain defect types based upon the inspection results.

In order to be able to focus testing based upon inspection results, knowledge about the relationships between inspections and testing is necessary. If, for example, an inspection technique is able to find a certain kind of defect that a subsequent testing activity is also able to find, the results of the inspection could be used to focus testing on certain defect types. However, such knowledge is (a) very rare or even not available in the existing literature [20], and (b) often context-specific. In general, such knowledge can be gathered by starting with observations that lead to assumptions, which are further evaluated in a given context and adapted subsequently [21].

Earlier publications focused on the applicability of the In$^2$Test approach and its efficiency improvement potential [3], [4], as well as on a general framework for deriving context-specific knowledge [22]. With the In$^2$Test approach, effort for testing can be reduced or more defects can be found (in the above-mentioned studies, the effort reduction that could be achieved for testing was up to 34%). One major question is whether such inspection defect data are a suitable predictor of defect-prone parts. Therefore, we conducted two case studies, which first compared inspection with product metrics as defect predictors, and then evaluated different combinations of these metrics in order to evaluate their potential. Furthermore, this article provides a detailed process for gathering context-specific knowledge for the integrated inspection and testing approach (calibration) in a retrospective manner. This article extends an earlier publication [23] by offering a clearer distinction between the calibration and the application process of the integrated approach, extends the related work, reports on feedback received from practitioners based on a questionnaire, and summarizes the main lessons learned.

The remainder of this article is structured as follows: Section 2 presents related work. It starts with a short overview of existing combinations of static and dynamic quality assurance and presents approaches that are used for focusing testing. Section 3 introduces the In$^2$Test approach, and explains the calibration process of the In$^2$Test approach. Section 4 presents experiences from two case studies where inspection and product metrics were compared, and additional qualitative results from a questionnaire. Section 5 derives the main lessons learned. Finally, Section 6 concludes the article and gives an outlook on future work.

## 2. Related Work

### 2.1. Combination of static and dynamic quality assurance

The combination of static and dynamic quality assurance techniques has received increased attention in recent years [2]. The objectives of such combinations are, for example, improved effectiveness, efficiency, or coverage of the system under test. Two main combination categories can be distinguished: compilation and integration approaches [2].

Compilation approaches comprise approaches that apply both static and dynamic analyses to improve a certain quality property, but without using results from each other. For instance, Zimmermann and Kiniry propose a combination of a static checker, a runtime assertion checker, and a unit-test generator [9]; Aggarwal and Jalote propose the combined application of a constraint solver and dynamic analyses [10]. On the other hand, integration means that one quality assurance technique uses the input from a second quality assurance technique to reduce the disadvantages of using them in a compiled manner. For instance, Godefroid et al. integrate symbolic execution, testing, and runtime analysis [11], and Chen et al. integrate model checking and model-based testing [12]. Furthermore, approaches explicitly integrating inspections and test-case generation, such as inspection-based testing (e.g., UBT-i, [13]), offer another way to integrate static and dynamic quality assurance techniques. However, such approaches are very rare and did not have the goal of focusing testing activities. We refer to [2] for a comprehensive overview and more details.

### 2.2. Focusing and Predictions based on Metrics

One well-established approach for focusing quality assurance activities is the prediction of defect distributions based on metrics. One of the earliest metrics, the famous cyclomatic complexity, was introduced by McCabe in 1976; thereafter, a plethora of metrics have been proposed, analyzed, and evaluated.

A major split is the differentiation into process and product metrics (the latter can be further distinguished). The first subcategory for product metrics is source code metrics. These metrics describe general characteristics of source code and can be applied to any programming language, e.g., class length measured in lines of code, or size measured in function points or number of methods. The second subcategory consists of object-oriented metrics. Strictly speaking, these metrics are also source code metrics. However, they make explicit use of object-oriented concepts such as generalization or specialization and therefore cannot be applied to all programming languages. A typical example of an object-oriented metric is 'inheritance coupling', which measures the number of parent classes to which a given class is coupled. The third subcategory, consisting of six further object-oriented metrics, is the well-established metric set introduced by Chidamber and Kemerer called the CK metric set [5]. The CK metric set consists solely of object-oriented metrics; however, a distinction between the CK metric set and other object-oriented metrics is reasonable, as the CK metric set has been empirically validated many times by various researchers in different contexts. Some examples of process metrics are number of revisions, number of authors, or number of refactorings. More details from a literature survey done by Kremer can be found in [24].

Certain metrics are more prominent in the literature than others. In order to get an idea of the practical potential of individual metrics, we analyzed how often they had been empirically validated in the 28 articles found. Figure 1 shows an overview of product and process metrics that were evaluated at least six times in the set of articles found.

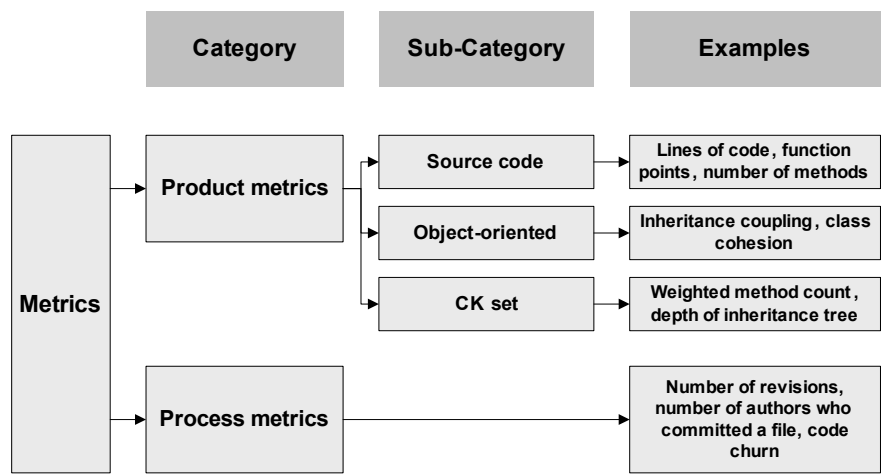

Fig. 1. Classification of metrics

Even though metrics such as 'lines of code' are evaluated often, the results vary significantly. For instance, Gyimóthy et al. stated that large classes are more defect-prone [7], whereas Fenton and Ohlsson conclude that smaller classes are not less likely to be defect-prone than larger classes [8]. One main difference between the two case studies is the different contexts they were conducted in. Consequently, there is no single set of metrics that fits all project contexts [6], and a metric set that fits best in a new context has to be identified before it can be applied to conduct predictions. D'Ambros et al. summarized many studies and defined an evaluation framework to assess different metrics more thoroughly [14]. Besides classic metrics such as size and complexity, the authors also analyzed more complex metrics such as entropy of changes, code churn, or combinations of historical and product metrics. Another comprehensive analysis of defect-prediction approaches is given by Arisholm et al. [15], who again showed that the performance of metrics with respect to their defect-proneness prediction depends on the context. However, inspection metrics are neither considered for the prediction of defect-proneness nor do further studies exist about their relationships with defect distributions.

Besides focusing testing based upon different kinds of data (which is the focus of this article), further approaches for focusing testing include the consideration of expert knowledge, the use of additional comprehensive output from static analysis tools, or the use of defect type classifications.

## 3. Integrating Inspections and Testing

### 3.1. The In²Test Approach at a Glance

As shown above, we identified a lack of approaches that take inspection defect data into account to focus testing activities. It makes sense to integrate these two quality assurance techniques in order to exploit synergy effects as they have the same objective, namely finding defects. Therefore, we developed the integrated inspection and testing approach In²Test [3], [4]. In²Test integrates inspections with testing techniques, i.e., inspection defect data is explicitly used to predict defect-prone parts in order to focus testing activities on those parts. Furthermore, In²Test is also able to focus on defect types; however, this is not the focus of this article.

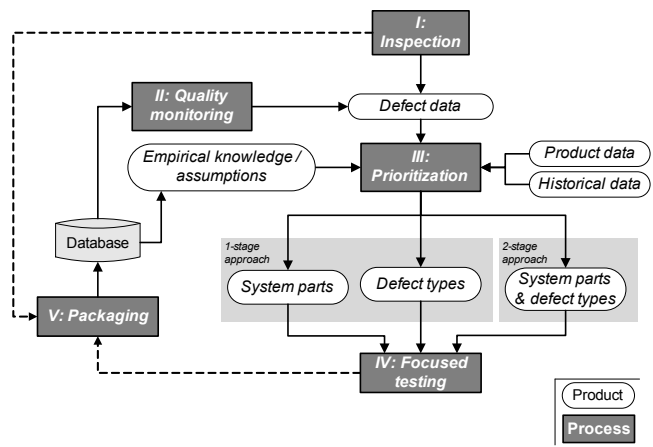

Fig. 2. In²Test approach at a glance

An overview of the approach, its process steps, and the information flow is given in Figure 2. First of all, an inspection is performed (**I**). The approach does not impose any prerequisites regarding a specific inspection technique in order to remain flexible, i.e., very formal inspection techniques such as Fagan inspections are possible, as are reviews, walkthroughs, or desk checks [25]. However, defect data from the inspection is expected, and different metrics can be gathered, such as number of defects per module or defect density per code class. Because inspections are a manual quality assurance technique and the performance of an inspection is dependent on certain context factors, such as inspector experience or process conformance, the quality of the inspection results is monitored (**II**) to make the results reliable. For this purpose, additional information such as the reading rate of an inspection can be gathered and analyzed in addition to the defect data. Further product data (e.g., size in lines of code, or number of developers per code module) or historical data can be considered in addition and could be combined with inspection metrics to improve prioritization. This is followed by prioritization of either system parts or defect types alone (1-stage approach), or of both (2-stage approach; i.e., system parts such as code classes are prioritized, and in those code classes, only certain defect types are addressed) (**III**). In an ideal **application** case, prioritization is done on the basis of existing context-specific empirical knowledge. If we know, for instance, that if many inspection defects are found in one area, more defects are expected during testing in this area, testing can be focused on those parts based on the currently gathered inspection metrics. Once prioritization is completed, a focused testing activity is performed (**IV**) using, for example, existing test cases or defining new ones (e.g., on the unit level, equivalence partitioning or boundary value analysis might be used). The data from the inspection and testing activity should also be stored in a database for further analysis (packaging, **V**).

If we return our attention to prioritization, we find that context-specific knowledge about the relationships between inspections and testing is often not available. In this case, the In$^2$Test approach has to be **calibrated** first, i.e., assumptions have to be stated and evaluated before the gathered knowledge can be used to predict defect-prone parts. The remainder of this article focuses only on how to perform the calibration and how the approach is adapted in this case, and presents experiences made with the calibration.

### *3.2. Calibration of the In$^2$Test Approach*

Before the In$^2$Test approach can be applied as shown in [3] to predict defect-prone parts or defect types based on inspection results, the calibration has to be performed for the new context, i.e., context-specific knowledge about the relationships between inspections and testing has to be gathered and evaluated (optionally, further data may be considered, such as product data). To do so, assumptions have to be defined. In order to be able to perform an objective evaluation of such assumptions, inspection and test data have to be gathered first. Thus, inspections and testing are conducted without influencing each other, i.e., inspection results are initially not used to focus testing. Rather, a traditional (i.e., unfocused) testing activity is done without the use of insights from the inspection. The concrete steps in the calibration of the In$^2$Test approach are (each step will be described in detail below):

  A. Execution of inspection and testing
  B. Prioritization
  C. Evaluation
  D. Packaging

*3.2.1.Execution of inspection and testing*

In order to be able to gather knowledge about the relationships between inspections and testing in a certain context, defect data from these activities have to be considered. If historical defect data from inspections and testing are available, this can be used as starting point. If not, the two quality assurance techniques have to be applied during software development and defect data have to be elicited. Reasonable metrics are, for example, number of defects per module, defect density per module, defect severity, or defect type. The defects found during the inspection need to be corrected before testing is conducted.

To ensure that the inspection data is suitable for predicting the remaining defect-prone parts, the quality of the inspection results has to be monitored. If inspections were already performed in that environment and it is known what the reading rate or the average defect density in the given context is, such data can be used for comparison and for judging the suitability of the current inspection results. If such data is not available, data from the literature can be used for comparison (e.g., [16]).

*3.2.2.Prioritization*

Before the prioritization starts, the context has to be gathered as the behavior between inspections and testing is usually context-specific. This means that those context factors have to be considered that have an influence on the defect detection of inspections and testing, and on defect injection. Some examples of such factors are the experiences of the inspectors and testers, the maturity of the software being checked, the development process, or a specific domain. A large set of possible influence factors can be found in [17]. The In$^2$Test approach currently considers context factors only informally to characterize the environment, which allows explaining, e.g., defect numbers.

In case no explicit and reliable context-specific knowledge about the relationships between inspections and testing is available, the prioritization step continues with the definition of new assumptions. Assumptions can basically be derived in two different ways: **analytically** or **empirically**.

Analytically means that based on a systematic analysis of a certain environment, which includes the consideration of process and product structures (e.g., development and quality assurance processes; experience of developers, inspectors, and testers; size and complexity of product to be developed), assumptions regarding the relationships can be derived in a logical manner. One example is: "It is assumed that a significant number of defects still remain in those parts that have not been inspected. Consequently, testing should focus particularly on those uninspected parts of a system." One rationale is that due to missing resources or time constraints, some parts might not have been inspected. In order to check the quality of those parts, the inspection results can be used to focus testing on different parts.

Empirically means that based upon empirical knowledge from different environments and new experiences from a given context, assumptions regarding relationships can be derived via data. First, accepted empirical knowledge from different contexts can be used

*Frank Elberzhager, Stephan Kremer, Jurgen Munch, Danilo Assmann*

and adapted to a given context, then it has to be checked. Second, when performing certain processes in a given context, new observations may be made, resulting in new or refined assumptions. This means that new empirical knowledge about certain relationships is gained. One empirically validated assumption about defect distributions is the Pareto distribution, which can be adapted for the In$^2$Test approach in the following way:

> Assumption 1: Parts of a system where a large number of inspection defects are found indicate more defects to be found with testing.

Such an assumption is very general, e.g., it has to be clarified what "parts of the system" and "number of inspection defects" mean in order to be operational. A refinement of an assumption results in a selection rule, which consists of an instruction and a condition. For example, the assumption mentioned above can be refined into the following selection rule:

> Selection rule 1.1: Select those code classes where the defect content (dc) is high.

The assumption is instantiated for the code level, and an absolute metric is used to express inspection defects. In order to judge the validity of a selection rule, a significance value can be added. Here, a validity of zero means that the selection rule is applied successfully zero times in the given environment (we will explain what successfully means in the Evaluation Section C). Each new selection rule gets a zero.

In order to be able to prioritize concrete code classes for testing, it has to be clarified next what "high" means in the given context. For this, the inspection defect data is considered, and an 80% rule could be applied for instance, meaning that the code class with the highest defect content is taken and the 80% value of that number is calculated and set as the threshold that has to be reached by code classes in order to be selected for testing. If, for example, the highest defect content of a number of inspected code classes is 14, the threshold would be set to 11, and selection rule 1.1 is refined into:

> Selection rule 1.1: Select those code classes where the defect content (dc) is higher than 11.

After the threshold is defined, prioritization can be done easily. Currently, the In$^2$Test approach would focus on the selected code classes during testing and omit the non-prioritized ones.

Figure 3 shows the five described steps during prioritization and the already stated example. In addition, a second example for a selection rule is shown. This selection rule considers a relative defect value, namely defect density, which is defined by the number of defects divided by the length of a code class. The threshold is calculated using a 50% value this time. In the example, four code classes are inspected. The application of the two selection rules results in two different prioritizations, which have to be evaluated.

The prioritization scales in a way that allows defining a lot of different assumptions and deriving a lot of specific selection rules for each assumption.

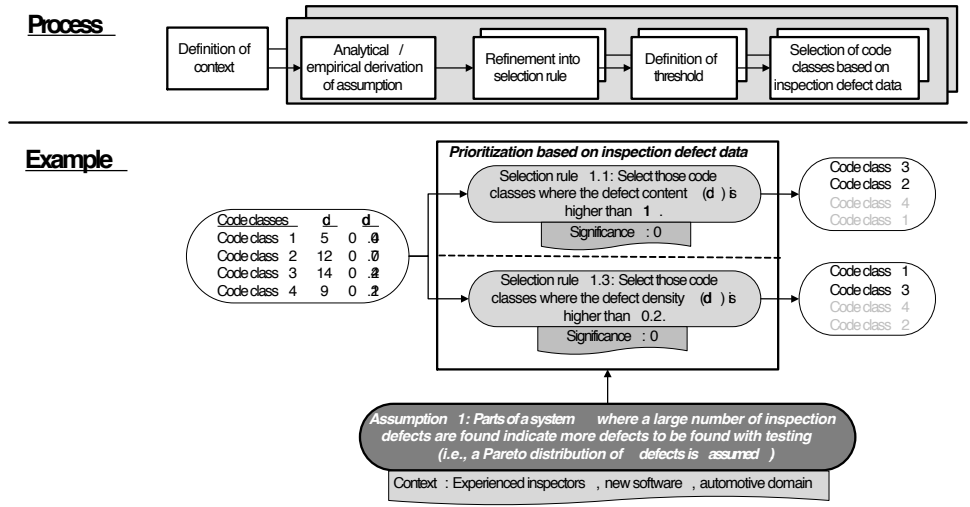

Fig. 3. Prioritization procedure including code level example

### 3.2.3. Evaluation

Besides the available inspection defect data, all test defect data are already available. Therefore, the evaluation can be performed directly after the prioritization. As already sketched during the prioritization step, assumptions may be contradictory, and it has to be checked which ones are valid in a given context. We assume that the context is stable and was gathered correctly.

Consequently, we distinguish two different cases during the evaluation: A selection rule and its corresponding assumption either led to a correct prediction or they did not. This means that the main criterion for being classified as a successful selection rule is whether the evaluation shows that all defects were found by the prioritization (this is called effective), or whether defects were overlooked (i.e., ineffective). The two cases can be further refined as shown in Figure 4.

Four selection rules are shown that use the already mentioned Pareto distribution, and selection rules 1.1 and 1.3 are already shown during the prioritization step. Two additional selection rules are defined.

Again, the first selection rule focuses testing on those code classes where the inspection defect content was higher than 11. Two code classes fulfill this condition, and are therefore prioritized for testing. During testing, more defects are found in only these two code classes (third column), i.e., the selection turned out to be ideal. The second selection rule, which uses a 50% criterion for defining the threshold, selects three code classes based upon the inspection results. After testing, it turned out that all defect-prone code classes were selected, but one additional code class was selected that was not defect-prone, which decreases efficiency. Therefore, the quality category is two. The significance of both selection rules 1.1 and 1.2 would be increased by 1.

| Assumption and selection rules | Inspection defects and prioritized code classes | Test defects | Evaluation rules & quality categories | |
|---|---|---|---|---|
| *Assumption 1*: Pareto distribution of defects is assumed. | | | | *effective* |
| *Selection rule 1.1:* defect content > 1 | Code classes   d<br>Code class 1   5<br>**Code class 2   2**<br>**Code class 3   4**<br>**Code class 4   9** | Code classes   d<br>Code class 1   0<br>Code class 2   8<br>Code class 3   1<br>Code class 4   0 | All parts in which test defects are prioritized, and parts in which no test defects are found are not prioritized. | 1 |
| *Selection rule 1.2:* defect content > 7 | Code classes   d<br>Code class 1   5<br>**Code class 2   2**<br>**Code class 3   4**<br>**Code class 4   9** | Code classes   d<br>Code class 1   0<br>Code class 2   8<br>Code class 3   1<br>Code class 4   0 | All parts in which test defects are found are prioritized, but also parts in which no test defects are found are prioritized. | 2 |
| | | | | *ineffective* |
| *Selection rule 1.3:* defect density > 0.2 | Code classes   d<br>**Code class 1   0.8**<br>Code class 2   0.0<br>**Code class 3   0.2**<br>Code class 4   0.2 | Code classes   d<br>Code class 1   0.0<br>Code class 2   0.2<br>Code class 3   0.5<br>Code class 4   0.0 | Only some parts in which test defects are found are prioritized. | 3 |
| *Selection rule 1.4:* defect density (crit. defects) > 0.2 | Code classes   d<br>**Code class 1   0.2**<br>Code class 2   0.0<br>Code class 3   0.8<br>Code class 4   0.0 | Code classes   d<br>Code class 1   0.0<br>Code class 2   0.3<br>Code class 3   0.2<br>Code class 4   0.0 | No part which is defect-prone is prioritized. | 4 |

Fig. 4. Quality of selection rules

Selection rule three selected only some of the defect-prone code classes, i.e., defects were overlooked during testing when only the prioritized code classes were selected; and selection rule four did not select any of the defect-prone code classes. Both are classified as ineffective and are classified into quality category three, respectively four. The significance of both selection rules 1.3 and 1.4 would remain zero.

With this classification, an easy-to-calculate overview of the suitability of each selection rule can be obtained, and it is possible to assess the effectiveness and efficiency of each rule. Of course, selection rules classified as effective, respectively those with a high significance value, should get special consideration for prioritization in subsequent quality assurance runs. Selection rules that were ineffective have to be treated with caution, and most often can be neglected.

The more quality assurance runs are conducted and the more selection rules and assumptions are evaluated in a context, the better the relationships between inspections and testing regarding defect-proneness distributions are known. One representation of a so-called trend analysis with respect to the introduced classification can be found in Figure 5. The quality categories of four different selection rules over four quality assurance runs are shown. Selection rules 1.1 and 1.3 both have a significance of four, but the former is slightly better in terms of efficiency. Both are reasonable candidates for subsequent prioritizations. Selection rule 1.2 was effective in two quality assurance runs, and ineffective in another two. The resulting significance is two; this selection rule could be analyzed in more detail to explain why this selection rule became worse. One reason might be that an influence factor changed significantly. The last selection rule showed bad results and should not be considered for future quality assurance runs.

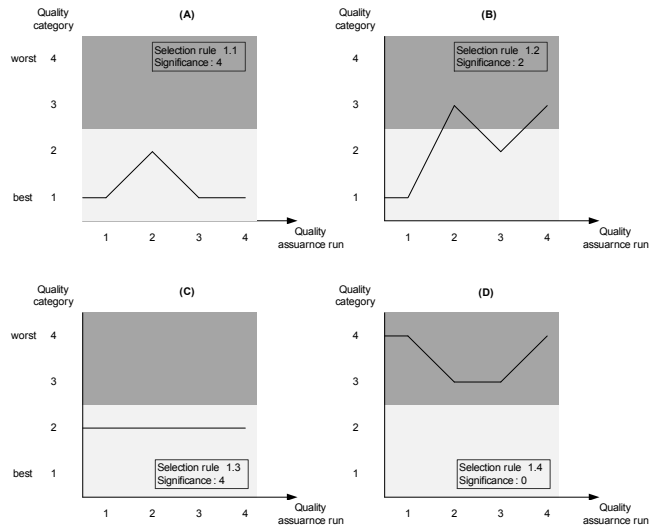

Fig. 5. Trend analysis of exemplary selection rules

Besides the introduced 4-category classification scheme, further evaluations might be considered. For example, the calculation of precision and recall could provide a more fine-grained assessment of each selection rule [15], and a calculation of average values using, e.g., the F-measure might also provide a good overview in case more rules have to be compared. Furthermore, regression models could be derived and correlations could be calculated [6].

### 3.2.4. Packaging

After the evaluation has been completed, all results, the defect data, and the gathered context information are stored in a database so that they can be reused for further analysis during subsequent quality assurance runs or for using the gathered knowledge for predictions of defect-proneness to focus testing based upon inspection results. If the number of data to be analyzed is manageable, alternatives like spreadsheet or statistical tools may be used for data storage in addition to the data analysis.

### 3.2.5. Summary

Figure 6 summarizes the four steps (A-D) during the calibration of the approach.

First, inspection and testing are performed, and the inspection results are monitored in terms of quality. The inspection defect data (e.g., defect content, defect density) and optionally further data (e.g., size, complexity) are used during the prioritization, followed by the evaluation of the assumptions and the selection rules considering the test results. New selection rules and assumptions might be defined and evaluated accordingly, i.e., the prioritization and the evaluation step can be performed multiple times. Afterwards, the results are packaged and can be reused in the next quality assurance run.

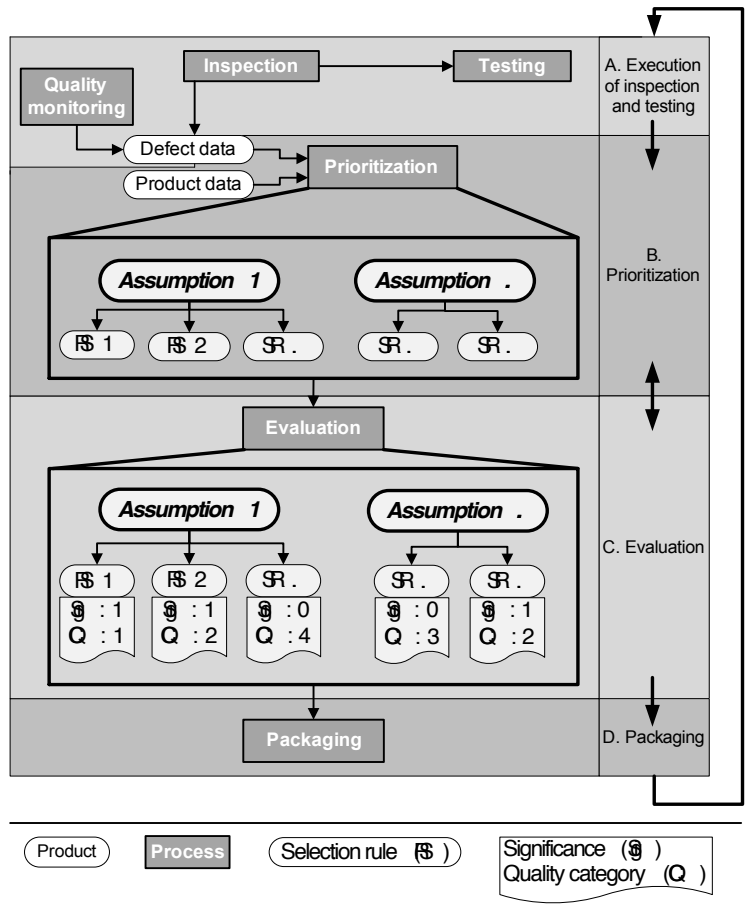

Fig. 6. Four steps when calibrating the approach

**4. Empirical Studies**

We conducted two kinds of empirical studies to further investigate the In²Test approach. On the one hand, we were interested in qualitative feedback about the expected performance of the approach. For this, we prepared a questionnaire, which was answered by practitioners after a presentation of the approach. On the other hand, we were interested in quantitative results of the approach, especially in a benchmark of inspection and product metrics for the prediction of defect-proneness. We extended the analysis of two earlier studies in order to evaluate different assumptions that consider various inspection and product metrics for prioritizing defect-prone parts. Consequently, in this section, we describe experiences regarding the performance of various product and inspection metrics for optimizing testing from two different contexts that we obtained during the calibration of the In²Test approach. The first one builds upon the results from a previous case study and extends these analyses [4]. The second one introduces experiences from an industrial environment. Furthermore, the results of a questionnaire to evaluate the approach and the expected performance are presented.

*4.1. Questionnaire*

In addition to the In²Test approach and its calibration procedures being applied to compare product and inspection metrics, the approach was also presented to twelve practitioners from different small- and medium-sized enterprises in January 2012. We presented the approach similar to the description given in Section 2, showed initial empirical results about the performance of In²Test (i.e., effectiveness and efficiency improvements gained from different case studies [3], [4]), and gave initial hints about how to apply the approach. After the presentation of the approach, which took about one and a half hours, a questionnaire was filled out by the participants of this workshop. The questionnaire consisted of two sections: (1) questions regarding experience and knowledge of the participants based on standardized categories and (2) questions regarding the approach itself based on the following categories: performance expectancy, effort expectancy, attitude towards using technology, facilitating conditions, and behavioral intention to use the system. These three categories were taken from a standard questionnaire [13]. With a few questions, we asked about the general context (e.g., company size or organizational role), to which the approach should be applied, but did not use this information further in this paper.

Categories 1-5 mentioned above used a 3-point Likert scale (2-agree, 1-neither agree nor disagree, 0-disagree) and were interested in getting an overall general impression regarding the integrated approach.

Table 1: Experience / Knowledge of participants, and question categories with results

| Question | all | developer | researcher | QA engineer | manager |
|---|---|---|---|---|---|
| Knowledge about software inspections / reviews? | 2.50 | 2.00 | 4.00 | 2.00 | 2.00 |
| Knowledge about software testing? | 3.42 | 2.33 | 4.00 | 4.00 | 3.00 |
| Experience with software inspections / reviews? | 2.00 | 1.33 | 2.67 | 2.00 | 2.00 |
| Experience with software testing? | 3.25 | 2.33 | 3.33 | 4.00 | 3.00 |

*Average experience and knowledge of participants (1 – little, 5 – a lot)*

| Category (# questions) | all | developer | researcher | QA engineer | manager |
|---|---|---|---|---|---|
| Performance expectancy (4) | 1.33 | 1.33 | 1.08 | 1.00 | 0.50 |
| Effort expectance (4) | 1.02 | 1.33 | 1.08 | 1.00 | 0.50 |
| Attitude towards using technology (4) | 1.24 | 1.42 | 1.33 | 1.00 | 1.13 |
| Facilitating conditions (3) | 0.81 | 0.89 | 1.00 | 0.67 | 0.67 |
| Behavioral intention to use the system (3) | 0.53 | 1.11 | 0.33 | 0.25 | 0.50 |

*Average results per category per participant group (0 – disagree, 1 – partly, 2 – agree)*

Regarding the roles of the questionnaire participants, we had three developers, four quality assurance engineers, three people focusing on research in their organizations, and two persons from management. A 5-point scale was used to gather experience and knowledge. The results can be found in the upper part of Table 1. Their knowledge and experience with testing was generally higher than with inspections.

With respect to performance expectancy (e.g., approach is useful, approach decreases effort, approach improves quality assurance), the overall average rates were between 1.17 and 1.42 depending on the question, which is a slightly positive result. The developer and quality assurance group showed the highest values (up to 1.75 on average). "Using the In²Test approach is a good idea" got the highest overall rating with 1.83. However, "I would find the In²Test approach easy to use" got only a 0.75 on average, while "My interaction with the In²Test approach would be clear and understandable" got 1.25 on average. One explanation is that applying the approach seems rather clear, but generating the empirical context-specific knowledge and using it is rather difficult. That was our motivation to develop the procedures for the calibration of the approach as shown in Section 3. Finally, questions regarding potential application of the approach within the next six months were answered rather negatively, mainly due to inspections not being done yet.

The bottom part of Table 1 gives another overview of the average results of each category presented, and gives the values for different groups. The managers generally provided the lowest numbers. One solution for improving their feedback might be to apply the approach with respect to a small project in order to convince them of the benefits of the approach in their context. Developers provided the most positive answers as they might benefit most from an approach that reduces the number of defects (and thus, rework effort). Quality assurance engineers also expect a high benefit; however, the missing concrete application guidelines prevented them from giving a higher score. Furthermore, the pre-conditions needed for the application of the approach were seen skeptically by most of the participants. The main reason for that is that either no inspections or reviews were being conducted yet, or that they were conducted informally, with no data being gathered by these practitioners.

Obviously, these results provide only a rough idea by some practitioners about the relevance of the In$^2$Test approach based upon a presentation. However, the feedback was rather positive with a lack in intending a concrete implementation of In$^2$Test due to only few applications of inspections so far. Furthermore, it led to an improvement (i.e., a more detailed process) regarding the use of the approach when first applying it in a new environment where no knowledge about the relationships between inspections and testing exists.

### 4.2. Case Study 1

#### 4.2.1. Goal

The main goal of this study was to evaluate the performance of certain well-established product metrics and inspection defect metrics that are able to focus testing activities. In two earlier case studies [3], [4], the In$^2$Test approach had been evaluated with regard to its feasibility and effort improvement potential. For this, defect and effort numbers gathered during two quality assurance runs, during each of which inspection and testing activities were conducted, were analyzed and compared to each other. However, only a small number of assumptions and selection rules were applied in this study, and no explicit comparison of inspection defect metrics to pure product metrics was conducted. Therefore, in order to compare the integrated inspection and testing approach In$^2$Test with established approaches using product metrics for focusing testing activities, the following two research question were derived:

Research Question 1 (RQ1): Which assumptions and selection rules that consider various inspection and product metrics lead to the best prioritizations of defect-prone code classes?

Research Question 2 (RQ2): Which assumptions and selection rules that consider various inspection and product metrics are stable across several quality assurance runs in a given environment, i.e., which assumptions and selection rules turned out to be most effective during a trend analysis?

*4.2.2. Context*

A case study in which two quality assurance (QA) runs were conducted in the same environment in order to analyze the In$^2$Test approach formed the basis for the following analysis. The approach was applied on the unit level.

The artifact to be checked was a Java prototype tool called JSeq, which had mainly been developed by one developer. JSeq supports practitioners in performing sequence-based specifications. At the time of the case study, it consisted of 76 classes, over 650 methods, and about 8,500 lines of code (LoC). The critical code parts were inspected. In the first QA run, these comprised four classes with a total of about 1,000 LoC. In the second run, four classes of about 2,400 LoC were inspected. Due to continuous development of the tool, the inspected code classes were completely different between the two QA runs.

In the first run, one inspector had very good inspection knowledge, but only limited programming experience, whereas the remaining three inspectors were mainly testers or developers with some inspection knowledge, but high programming experience. In the second run, one developer was replaced by an experienced inspector. The testing activity was performed by one developer who was not involved in the inspection.

In the first run, only one assumption considering inspection defect data was applied in order to check the general applicability of the integrated approach. Three assumptions applied in the second run considered only inspection defect data, the combination of inspection defect data and size, and the combination of inspection defect data and complexity. These initially stated assumptions were defined in a group session, and general empirical evidence for each one was found in the literature. A set of derived selection rules that were analyzed in the second run showed an effort improvement for test execution (including test definition) of between 6 and 34 percent (we refer to [4] for more details; the actual selection rules are shown in Section 4.2.4).

*4.2.3. Design*

In order to perform a detailed analysis of product and inspection metrics for focusing testing activities, we used data from the original study, which had analyzed the In$^2$Test approach during two QA runs [4]. In the study, a code inspection was conducted first each time, followed by quality monitoring of the inspection results, prioritization of the code classes based on the inspection results and further metrics, and application of a unit test. Afterwards, a retrospective analysis of the suitability of the initially defined assumptions and selection rules was conducted. The analysis in this study considers assumptions and selection rules focusing on defect-prone code classes, i.e., no prioritization of defect types is considered here.

In order to be able to perform a comprehensive comparison of different inspection and product metrics and their combinations, the initial set of assumptions and selection rules was heavily extended (i.e., tripled) in a systematic manner. Furthermore, an evaluation scheme had to be selected. For this, we used both a broad- and a fine-grained scale as shown in Section 3, i.e., first each selection rule was assessed as being either effective or ineffective. Besides this coarse-grained assessment, the more fine-grained one with four quality categories was used.

Besides the individual analysis of assumptions and selection rules during each of the two QA runs, a trend analysis was performed in order to check which of the selection rules were suited best across both QA runs.

*4.2.4. Execution*

In the original study [4], only three assumptions, respectively 32 selection rules were used during the evaluations. The objective in this article is to provide a more comprehensive evaluation of assumptions and selection rules that are valid in the given context, with explicit comparison of inspection and product metrics. All necessary information had already been gathered during the two QA runs (e.g., defect information for each code class, two size and one complexity metrics).

In this study, we defined fourteen different assumptions based on empirical knowledge available in the literature. Two of the assumptions only consider inspection metrics, four consider product metrics, and eight combine inspection metrics and product metrics. For example, with respect to inspection metrics, the following two assumptions are reasonable:

- A.I.: Parts of the code where a large number of inspection defects are found indicate more defects to be found with testing.
- A.II.: Parts of the code where a low number of inspection defects are found indicate more defects to be found with testing.

For each of these assumptions, detailed selection rules were derived systematically, resulting in an overall number of 118 selection rules. Table 2 shows the calculation and the actual assumptions and selection rules. Considering assumption A1 again, the following eight selection rules were derived:

- Focus testing on those code classes with large defect content considering all inspection defects.
- Focus testing on those code classes with large defect content considering high-severity inspection defects.
- Focus testing on those code classes with large defect content considering medium-severity inspection defects.
- Focus testing on those code classes with large defect content considering low-severity inspection defects.
- Focus testing on those code classes with large defect density considering all inspection defects.
- Focus testing on those code classes with large defect density considering high-severity inspection defects.

- Focus testing on those code classes with large defect density considering medium-severity inspection defects.
- Focus testing on those code classes with large defect density considering low-severity inspection defects.

Table 2: Calculation numbers of selection rules

| Assumptions | | Selection | Metrics one | Metrics two | # |
|---|---|---|---|---|---|
| I | inspection defect data | 2 x | 2 x | 4 | = 16 |
| II | | large / small | defect content / defect density | all defects / high severity defects / med. severity defects / low severity defects | |
| III | size | 2 x | 1 | | = 2 |
| | | large / small | class length | | |
| IV | size | 2 x | 1 | | = 2 |
| | | large / small | method length | | |
| V | complexity | 2 x | 1 | | = 2 |
| VI | | high / low | McCabe complexity | | |
| VII | inspection defect data + size | 4 x | 2 x | 4 | = 32 |
| VIII | | large + large / large + small / small + large / small + small | defect content + class length / defect density + class length | all defects + LoC / high severity defects + LoC / med. severity defects + LoC / low severity defects + LoC | |
| IX | inspection defect data + size | 4 x | 2 x | 4 | = 32 |
| X | | large + large / large + small / small + large / small + small | defect content + method length / defect density + method length | all defects + LoC / high severity defects + LoC / med. severity defects + LoC / low severity defects + LoC | |
| XI | inspection defect data + complexity | 4 x | 2 x | 4 | = 32 |
| XII | | large + high / large + low / small + high / small + low | defect content + McCabe / defect density + McCabe | all defects + McCabe / high severity defects + McCabe / med. severity defecty + McCabe / low severity defects + McCabe | |
| XIII | | | | | |
| XIV | | | | | |
| | | | | Sum: | 118 |

The same number of selection rules was derived for assumption A2, where it is assumed that a small number of inspection defects lead to more defects to be found during testing. These two assumptions result in 16 selection rules. The selection rules for assumptions A3 to A14 are calculated accordingly. Some further examples of actual selection rules are the following: For A.III: "Focus testing on large code classes." A final example of A.XII is: "Focus testing on code classes with large defect content and low complexity considering high severity inspection defects and low McCabe complexity."

In our context, the lines of code also consider blank and commentary lines. Defect density is defined as the number of defects divided by the lines of code.

A definition of the general selection criteria "low", "large", etc. was done context-specifically during the original case study, so that all of these selection rules could be applied with respect to the gathered data during the two QA runs in order to allow assessing and comparing them. To give an idea of the defect data and product metrics, Table 3 shows an excerpt (we refer to [4] for the remaining data). For example, the threshold for large number of defects in the second QA run was defined in our context as 32 (i.e., code class with highest defect number – 20%). This means that we once again followed the Pareto principle to define concrete values for our context. However, other rules might be used to calculate thresholds.

For analyzing and packaging the results, we used a spreadsheet tool. This allowed us to manage the defect and product data, and to easily evaluate which assumptions and selection rules were of the highest benefit.

Table 3: Excerpt of metrics

|  | QA run 1 |  |  |  | QA run 2 |  |  |  |
|---|---|---|---|---|---|---|---|---|
| **Code classes** | I | II | III | IV | V | VI | VII | VIII |
| *Inspection defects* | 26 | 6 | 27 | 8 | 14 | 40 | 39 | 7 |
| *Test defects* | 3 | 0 | 4 | 0 | 0 | 0 | 6 | 0 |
| *Class length* | 469 | 37 | 275 | 243 | 231 | 1364 | 701 | 115 |
| *Mean method length* | 4 | 9 | 7 | 177 | 3 | 14 | 8 | 7 |
| *McCabe complexity* | 2 | 5 | 2 | 44 | 2 | 4 | 3 | 2 |

*4.2.5.Results*

**QA run 1:** In order to answer RQ1, we first analyzed the 118 selection rules with respect to the first QA run. Nineteen selection rules turned out to be effective, and consequently, ninety-nine were ineffective. This is not surprising as we analyzed a large number of rules. Figure 4 gives an overview with respect to the four different categories.

In our context, the best selection rules (i.e., category A) were those that use large defect content alone or combine this with large class length, small method length, or low complexity. Thus, a Pareto distribution could be confirmed. Rules considering large class length or small method length led to category B. Defect density was a bad predictor for defect-proneness in our context (category C). This means that certain inspection metrics alone and combined with traditional product metrics led to the best selections of defect-prone code classes here, and product metrics alone led to suitable predictions but not to the most efficient ones.

In conclusion, assumptions considering large numbers of inspection defects and low complexity were appropriate. With respect to size, it depends on the concrete size metric. Corresponding combinations also led to suitable selections.

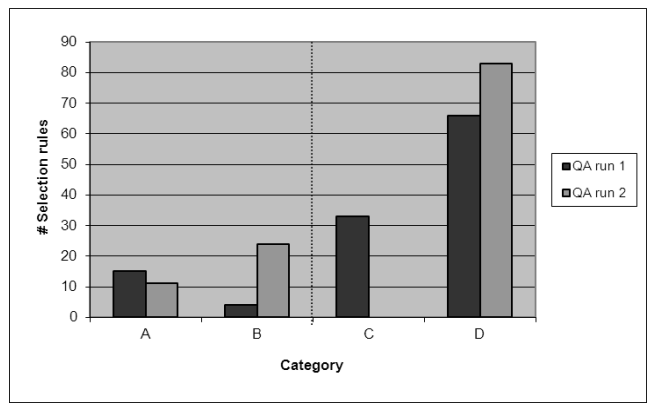

Fig. 7. Assessment result of selection rules

**QA run 2:** Next, we analyzed the 118 selection rules with respect to the second QA run. Thirty-five selection rules were rated as effective. The number for category B increased, which is not surprising due to the fact that only one defect-prone code class was found during testing, and many selection rules select more than one code class. No category C selection rule was found because no subset of one defect-prone code class can exist. However, the general trend of A+B and C+D is comparable to the first QA run.

Again, large defect content alone and the combination with large class length or small method length led to the best selections of code classes. However, large defect density led to much better results in the second QA run. Furthermore, high complexity alone and combined with large defect content and defect density led to suitable results (instead of low complexity as in the first QA run). Selection rules considering large class length or low method length were again evaluated as category B.

Consequently, the Pareto distribution could be confirmed again. While the two size metrics showed similar results compared to the first QA run, namely being effective predictors of defect-proneness while not being most efficient, complexity behaved inconsistently.

Moreover, a combination of inspection and product metrics for focusing testing activities showed the best results in our context (category A). A large number of selection rules were found that led to ineffective results and that are of little relevance for future QA runs in the given context.

In conclusion, assumptions considering large numbers of inspection defects and high complexity were appropriate. With respect to size, it depends on the concrete size metric again. Corresponding combinations of inspection and product metrics also led to suitable selections of code classes that were most defect-prone during testing, i.e., those code classes were prioritized by the selection rules for testing.

**Trend analysis:** In order to answer RQ2, we analyzed which selection rules were effective, respectively ineffective, with respect to both QA runs. Figure 8 presents an overview of the results.

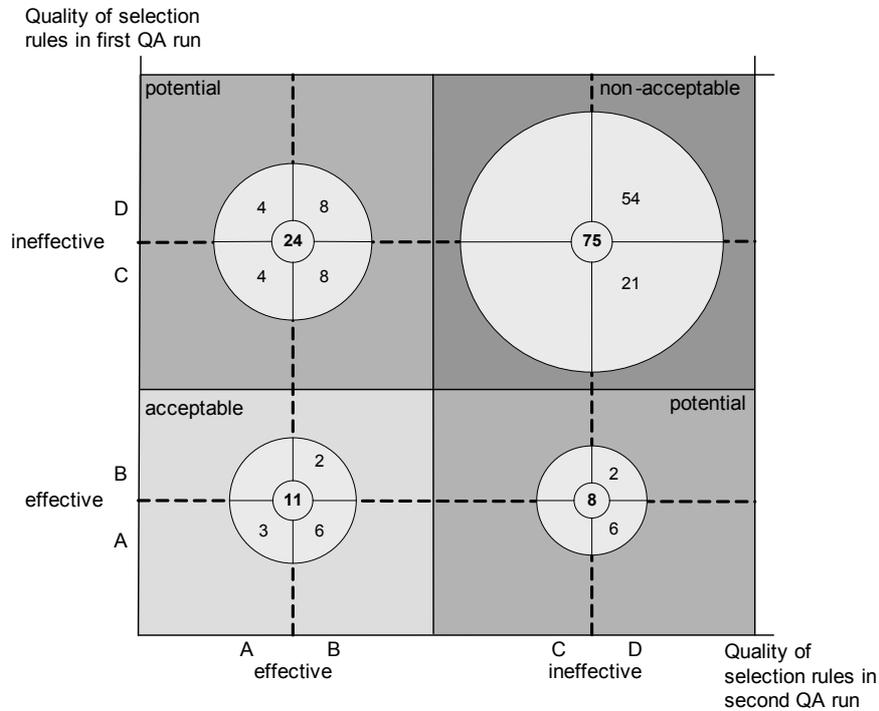

Fig. 8. Trend analysis of selection rules

First of all, the general classification of selection rules into effective and ineffective ones over the two QA runs revealed that only about 10% of the selection rules were effective in both runs (acceptable quadrant). These eleven rules are promising candidates in the given environment for a highly effective prediction of defect-prone parts. About 30% showed good results in one run, both bad in the other (potential quadrant). Those should be further analyzed, e.g., whether certain context factors can explain those differences, and how they behave in subsequent QA runs. The remaining 75 selection rules showed ineffective results in both QA runs and are thus of little interest for future runs (non-acceptable box). The high number of such classified rules is not surprising, as we compared a large number of selection rules.

With respect to the acceptable selection rules, three were classified into category A in both runs, i.e., these selection rules selected exactly the defect-prone code classes for testing based on large defect content (all, medium, and low severity) combined with small method length. Six more selection rules showed very promising results, considering only large defect content, and large defect content combined with large class length. Two more selection rules were twice categorized as 'B', namely large class length and small method length. This means that inspection defect data alone (in terms of defect content) and inspection defect data combined with certain product metrics led to the best prioritizations in our context. Furthermore, two size metrics led to appropriate selections of code classes, though not to the most efficient prioritizations. This also holds for the corresponding assumptions. Table 4 lists those metrics that led to the best prioritizations of code classes containing defects found during testing.

Table 4: Best metrics for prediction defect-proneness in the given context during two QA runs

| Quality | Metric |
|---|---|
| AA | High inspection defect content and low method length (all, low, and medium severity) |
| AB | High inspection defect content (all, low, and medium severity) |
|  | High inspection defect content and high class length (all, low, and medium severity) |
| BB | High class length |
|  | Low method length |

A lot of selection rules considering high defect density alone or in combination with the aforementioned product metrics led to category C prioritizations and might lead to more suitable results in future QA runs. Furthermore, selection rules using complexity led to inconsistent selections in our context. While in the first QA run, low complex code classes were more defect-prone, this changed in the second QA run, and high complex code classes tended to be more defect-prone. One explanation is that the first QA run was performed when the software was still not very complex, and thus, such parts also contained defects.

*4.2.6. Threats to Validity*

Conclusion validity: The presented results are only based on two QA runs in a given context and therefore, a lot more evaluations in the same and in different contexts are needed before they can be generalized. However, first positive trends could be identified indicating that inspection defect results lead to good prioritization of defect-prone parts for testing, and that a combination with established product metrics might even improve such a prioritization.

Construction validity: A set of assumptions and selection rules was derived systematically for our analysis. However, a lot of additional ones might exist, and a comparison with more product metrics would strengthen such an analysis.

Internal validity: The evaluation of selection rules might have been done differently by other QA engineers who might have defined thresholds for 'large' or 'small' in a different way. However, the thresholds were discussed in a team of quality assurance engineers to reduce this threat.

External validity: The software under inspection and test during the QA runs was rather small, and only a small set of code classes was considered for the prioritizations. Furthermore, the performance of the applied assumptions and selection rules is initially only valid in the given context and cannot be generalized, as each such rule has to be evaluated again in each new context. However, we presented an initial set of assumptions and selection rules that showed promising results and which are consistent with existing empirical knowledge (such as the Pareto distribution); thus, our rules might serve as a starting point for evaluations in new environments, and can be enhanced by additional or alternative rules.

### *4.3. Case Study 2*

#### *4.3.1. Goal*

Based on the experiences from the first case study, we started an evaluation of the In$^2$Test approach in an industrial environment. The main goal of the second case study was to evaluate which assumptions and selection rules lead to the most effective prioritizations of parts for testing. Once again, different inspection and product metrics used to select defect-prone parts were compared. Similar to RQ1, RQ3 is stated as follows:

Research Question 3 (RQ3): Which assumptions and selection that consider various inspection and product metrics rules lead to the best prioritizations of defect-prone modules?

#### *4.3.2. Context*

The analyzed organizational unit has been developing software for embedded systems, mainly automotive, for over 20 years. Currently, Vector uses a product family approach for development with three levels of variation: the product as the full superset of all features, the program as the first derived level of variation for a specific customer platform, and the delivery as second level of variation for a concrete microprocessor and compiler. The features are implemented in the form of components.

As a consequence of the sensitive context, several activities are performed to ensure the quality of the software. On the code level, code inspections are conducted on all released source code. Testing is done on all elements of the product family: the product itself, programs, deliveries, and components. The defect data and all feedback from customers are stored in the form of change requests. Even though the development model of the product family has changed over time, test and inspection data are available spanning more than ten years.

The software currently consists of about 140 modules, each comprising a set of code classes. The size of a module varies between 120 and 14,000 statement lines of code.

#### *4.3.3. Design*

In order to evaluate the In$^2$Test approach and a set of different assumptions and selection rules, a retrospective design was chosen again. For this, existing inspection data had to be collected first, as all such data were documented across several change request documents. For our analysis, we concentrated on a subset of twelve of the available modules (about 10%), which we chose randomly. The reason is that size, complexity, age, and number of deliveries are highly heterogeneous. The first question was whether we could find basic assumptions that work independent of any component classification. For each module, a different number of change requests (and thus inspection data) existed. The oldest defect data that was considered was from 2007.

With respect to test data, we considered all defects found during various kinds of testing performed subsequent to the code inspection.

Furthermore, two code metrics (i.e., product metrics) were considered and calculated for the corresponding modules: size in lines of code and waste per source code line, which express the stability and sustainability of the developed code. A high value implies that many parts are changed or thrown away over time.

For the analysis of the approach, we defined ten assumptions, based upon the available data from the context and our experience from the first case study. Furthermore, due to the defect distribution, a categorization of assumptions into effective and ineffective turned out not to be useful because each module contained at least some minor problems over the considered timeframe, and all assumptions would therefore have been classified into the detailed category C (which would have made them impossible to compare). Therefore, we defined four selection rules for each assumption that, based on the general assumption, select, e.g. the three most defect-prone modules based on the inspection results, respectively the three largest modules based on the product metrics. We continued this with the top-5, the top-8, and the top-10 modules, and evaluated how many test defects had been found by such selections, i.e., we derived a sorted list instead of defining a set based on a hard threshold. With such an analysis, a baseline of appropriate assumptions could be gathered for the given context.

### 4.3.4. Execution

We first presented the In$^2$Test approach to Vector and discussed the expected benefit in their environment. As improving quality assurance is a major goal for Vector, we decided to evaluate the approach in a retrospective design based on the historical inspection and test defect data. We first gathered inspection data from several change request documents for the randomly chosen code classes, which took several days. The test data could be extracted easily from a defect tracking system. However, chronology sequence and relations between the inspection and test data were difficult to extract. Therefore, we decided to use the existing data to draw a baseline from which we assessed each assumption.

Ten assumptions were derived: four considering inspection metrics, four considering product metrics, and two combining inspection and product metrics. With respect to those using inspection defect data, a Pareto distribution was assumed, and different representations of the inspection data were used:

1. All inspection defect data for a module.
2. Like 1, but scaled (not all modules were inspected 100%, but the rate was given and could be used to estimate the inspection defect numbers if 100% had been inspected).
3. All inspection defect data without counting inspection comments.
4. Like 3, but scaled.

With respect to assumptions considering product metrics, modules of small and large size, and modules of small and large waste per line are taken.

Four selection rules were derived for each of the assumptions following the structure shown in Figure 9. One concrete example: "A1: Focus testing on the top-3 defect-prone code classes based on all inspection defect data."

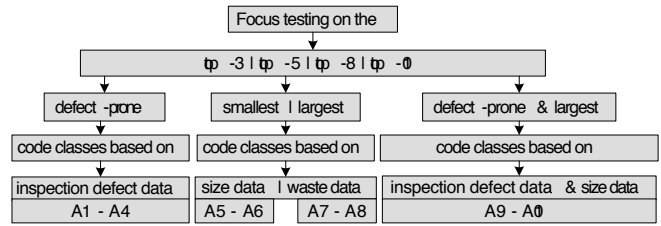

Fig. 9. Structure of applied selection rules

Based on the available defect data and the product metrics, all assumptions and selection rules were applied in a retrospective manner in order to evaluate their validity.

For analyzing and packaging the results, we again used a spreadsheet tool. This allowed us to manage the defect and product data, and to easily evaluate which assumptions and selection rules were of the highest benefit.

*4.3.5. Results*

Figure 10 presents an overview of the number of found test defects with respect to the ten assumptions. For each assumption, four different selection rules were evaluated. For example, assumption A1 assumes that a large number of the test defects were found in those modules where most inspection defects had been found before. With respect to the first selection rule, which considers the top-3 defect-prone inspection modules, about 30% of all test defects had been found (black bar in Figure 10). Considering the top-5 modules, more than 80% of all test defects had been found (black bar plus dark gray bar in Figure 10). This means that considering about 40% of all defect-prone modules based on the inspection defect data was sufficient for finding more than 80% of all defects during testing. Focusing on the top-8 defect-prone modules during inspections, more than 90% of all test defects were found. The top-10 do not further improve prioritization when using assumption A1.

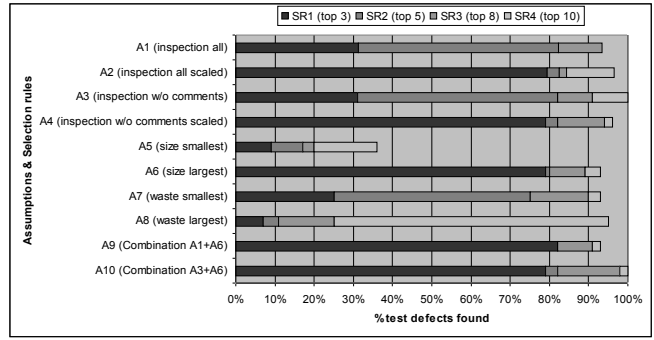

Fig. 10.  Results of different assumptions with respect to the percentage of defects found during testing

Without initially considering the combined assumptions A9 and A10, assumptions A2, A4, and A6 led to the best results with respect to the top-3 modules, i.e., those selection rules need only 25% of the defect-prone modules to focus testing in such a way that about 80% of the remaining defects are found. Two of these assumptions consider inspection results, one assumption considers size. With respect to the top-5, all four inspection assumptions led to suitable predictions (more than 80% of test defects were found). Only one selection rule (from assumption A3) revealed all defect-prone modules with the top-10 selection (and thus, can be classified as "effective" in the sense of the first case study). However, almost no selection rule was able to prioritize all modules containing defects in that context due to the fact that almost every module contained defects, and the selection should identify a subset of modules in order to save effort while maintaining the same quality. Selection rules considering small size and large waste did not lead to suitable results.

Finally, we combined inspection and product metrics and also calculated the number of defects found. For example, when combining A1 and A6 and calculating the top-3 value, we counted each module selected by one of the two assumptions, which resulted in two modules that were selected by either A1 and A6, and two modules that were selected by only one of the two assumptions. The resulting effectiveness value was 82%. Surprisingly, the top-5 calculation did not improve the value significantly, but the modules selected for the top-8 did. This indicates that one should focus on modules that fulfill the combined assumption best (i.e., top-3 focus) to find about 80% of the test defects, and on those modules in the top-5 to top-8 range of the combined assumption (by skipping those in the top-3 to top-5 range) to find another 15% of the remaining test defects in a given context.

Two main conclusions can be drawn: The assumptions that considered inspection defect data (1) led to suitable predictions, and (2) are of similar effectiveness as selection rules using size metrics. Though not all defects were found by most of the selections due to the long timeframe that was considered, the most critical parts could be identified by these assumptions. Because the modules were selected randomly, our objective is to further investigate whether these product metrics and inspection metrics behave similar with respect to a broader dataset, and whether the prediction can be further improved when combining them.

*4.3.6.Threats to validity*

Conclusion validity: The presented results are only based on the analysis of a subset of all available modules from the context. Therefore, a larger analysis is still necessary. However, the initial results substantiate similar trends from different environments.

Construction validity: It is possible to evaluate a set of additional assumptions and selection rules, which might lead to further conclusions. However, we chose such metrics in our assumptions and rules that were already available in the given context.

Internal validity: A certain degree of inaccuracy is often a fact with respect to historical defect data. However, the absolute number of documented defects was large enough to compensate for that to a certain extent.

External validity: First, it could be verified in the given context that large modules tend to be more defect-prone, which is consistent with evaluations from different contexts. Furthermore, the prioritization for testing based on the inspection metrics led to similar results for some assumptions, or even slight improvements compared to the product metrics. Though this is a positive trend in this environment, conclusions with respect to other environments have to be drawn with caution, as other assumptions might lead to good selections in different contexts. The assumptions used in this study can serve as a starting point for such evaluations.

**5.Lessons Learned**

Based on the experiences we made in the different environments, the essential lessons learned are:

- The In$^2$Test approach is relevant for quality assurance engineers in order to improve the overall quality of a product. Insights from inspections may give additional hints on how to focus testing activities.

- Inspection results can be a suitable predictor for defect-prone parts or defect types. The aim of the In$^2$Test approach is not to substitute existing prediction approaches, but to use inspection defect data as one additional metric during predictions to improve the overall prediction excellence. With such an approach, a more integrated and holistic quality assurance can be performed which exploits further synergy effects such as higher defect detection or lower efforts. Remembering that, for example, testing efforts in particular may consume up to 50% of the overall development effort, the integrated inspection and testing approach is able to significantly reduce such effort as already shown in [4].

- Selected product metrics led to good predictions for defect-proneness, similar to selected inspection metrics. Best results for defect predictions were achieved in our context when inspection and product metrics were combined.

- Gathering context-specific knowledge between inspections and testing (optionally: further metrics) is a prerequisite before using the approach for predicting defect-proneness. The calibration of the approach as shown in this paper was refined during different case studies, turned out to be applicable in an industrial context, and provides valuable insights for quality assurance.

- Assumptions and refined selection rules are context-specific. Their definition and evaluation are easily possible and the necessary effort is manageable, at least once first assumptions and selection rules have been defined and the basis for the evaluation has been created. However, the theoretical number of possible assumptions is tremendous, and a starting set has to be defined carefully. Besides inspection and test defect data, available product data might be a reasonable alternative for developing selection rules. Moreover, the metrics used in our case studies, which have to be further adapted and evaluated, could serve as valuable input. The first 4-scale evaluation scheme was sufficient for the initial context, but needed to be adapted for the industrial context.

- The prerequisites for applying the In$^2$Test approach are low, i.e., the approach does not require any specific inspection or testing technique, any specific development approach, or any specific domain. The minimal requirement is a set of inspection and test defects per system area. Historical inspection and test data may be a good starting point if available, but need to be carefully compiled, as the quality might be insufficient. However, a sound understanding and expertise with regard to data analysis may improve the validity of the results.

### 6. Summary and Outlook

Focusing testing by predicting defect-prone software parts is one way to improve quality assurance activities. Usually, product and process metrics are used for such a prioritization. However, inspection defect data is usually not considered. Therefore, we propose the In$^2$Test approach, which is able to use inspection defect data alone or in combination with such product and process metrics. A systematic 4-step process was introduced that can be used to gather the needed knowledge between the two quality assurance activities in a retrospective manner. This approach is not intended to substitute existing approaches, but rather complement them in order to further support the planning of quality assurance activities. We were able to show that the calibration of the approach is applicable, and new knowledge was gathered that helps to improve the overall quality assurance.

*Frank Elberzhager, Stephan Kremer, Jurgen Munch, Danilo Assmann*

In order to evaluate the suitability of inspection and product metrics for predictions of defect-prone parts, we conducted two studies and compared different metrics. We could show that inspection defect data were an appropriate predictor in those two contexts, which was further improved when inspection metrics were combined with certain product metrics. Our results represent promising, but initial results, and more empirical studies are necessary before generalizing our conclusions.

We are planning to continue evaluations of the In$^2$Test approach in order to substantiate our findings and to find more relationships between inspection and test defect data. Besides the calibration of the approach, a real application is one of the next steps, raising questions such as how to continuously evaluate selection rules during the application of the approach. Furthermore, a lot of additional product and process metrics exist that can be compared and combined with inspection defect data. Further evaluation of selection rules can substantiate our findings, e.g., by calculating precision and recall values. In addition, results from requirements and design inspections might help to better focus subsequent quality assurance activities. Finally, the approach could also be extended in such a way that test data might be used for improving the inspection, as the empirical concepts for evaluating assumptions and selection rules are generalizable.

**Acknowledgments**

This work has been funded by the Stiftung Rheinland-Pfalz für Innovation project "Qualitäts-KIT" (grant: 925) and the ARTEMIS project "MBAT" (grant: 269335). We would also like to thank Sonnhild Namingha for proofreading.